\documentstyle[11pt,paspconf,psfig,twoside]{article}
\begin{document}

\title{Indirect imaging of the accretion stream in HU Aquarii}

\author{M.K.~Harrop-Allin and M.~Cropper}
\affil{Mullard Space Science Laboratory, University College London, 
Holmbury St.~Mary, Dorking, Surrey, RH5 6NT, UK}
\author{P.J.~Hakala}
\affil{Observatory and Astrophysics Laboratory, Univ.~of Helsinki, Finland}
\author{C.~Hellier}
\affil{Department of Physics, Keele University, Staffordshire, UK}
\author{T.~Ramseyer}
\affil{Dept. of Physics and Astronomy, Univ. of Central Arkansas, USA}

\begin{abstract}
We apply our technique for indirect imaging of the accretion stream to {\em
UBVR} eclipse profiles of HU Aqr obtained when the system was in a high
state. We perform model fits to the eclipses using two geometries: one where
the stream accretes onto one footpoint of the field line, and the other where
the stream accretes onto both footpoints. We find that the stream brightness is
not uniform, nor is it a simple function of the radial distance from the white
dwarf. The model provides estimates of the radius where the stream couples to
the magnetic field ($1.0-1.3 \times 10^{10}$\,cm) and of the mass transfer rate
($8-76 \times 10^{16}$\,g/s). We examine the wavelength dependence of the
stream brightness, and find that in the case of the two-footpoint model, the
sections of the magnetically-constrained stream immediately adjacent to the
coupling radius appear to be hotter than the remainder of the stream. This may
be due to shock heating of the plasma as it threads onto the magnetic field.
\end{abstract}


\section{Scientific context}
The mechanisms whereby the accretion stream in polars becomes coupled to the
magnetic field are complex and not well-understood. In order to provide further
observational constraints to this problem, we have developed a method to image
the accretion stream in eclipsing polars. The technique has close parallels
with eclipse mapping of accretion discs in non-magnetic systems (e.g. Marsh \&
Horne 1988), in that both use photometric eclipse profiles to deduce the
distribution of emission between the component stars, and both make use of
maximum entropy regularization to constrain the problem. Our method operates by
placing emission points along a pre-set stream trajectory, and then
``observing'' the model stream through an eclipse by a Roche lobe-filling
secondary star. The relative brightnesses of the emission points are adjusted
to obtain the most locally smooth stream whose eclipse profile matches an
observed profile. The optimization is performed using a Genetic Algorithm in
order to maximize our chances of finding the global optimum in the
multi-dimensional space. For more details and tests of the method, see
Harrop-Allin, Hakala \& Cropper (1998). We present here the results of applying
this method to HU Aqr observed in a high accretion state ($V\sim 15.1$).
 
The data used for the modelling were obtained in August 1993 using the Stiening
photometer on the 2.1\,m reflector at McDonald Observatory. The imaging
procedure was applied to a total of five eclipses obtained over three
consecutive nights. The eclipse profiles are asymmetrical and show two chief
components: a steep component due to the accretion region on the white dwarf,
and a more gradual component due to the accretion stream. The stream's
contribution to the total emission is comparable, and in $U$ and $B$ {\em
exceeds}, the emission from the accretion region. There are prominent
pre-eclipse absorption dips at orbital phase $\phi \sim 0.88$ in all four
wavebands in the high state. The phase of the dip centre moves to successively
earlier phases during the course of the observations (which cover three
nights), implying an increase in the angle between the absorbing material and
the line of centres from $42^{\circ}$ to $ 48^{\circ}$. This dramatic movement
of the stream occurs without any change in the overall brightness of the
system.

\section{The stream brightness distributions}
Before the imaging procedure can be applied, a choice has to be made whether to
use a model stream that accretes at one pole or both. HU Aqr shows evidence for
both one and two-pole accretion in the high state. On one hand, the overall
shape of the light curve is consistent with cyclotron beaming from a single
pole. On the other hand, circular polarimetry obtained in June 1993 when the
system was also in a high state (Hakala, unpublished data) shows clear positive
and negative excursions, suggesting accretion onto (at least) two regions of
the opposite polarity. We therefore perform model fits to the eclipse profiles
using both geometries, and present sample results of each.

\begin{figure}
\hspace*{-2mm}\psfig{file=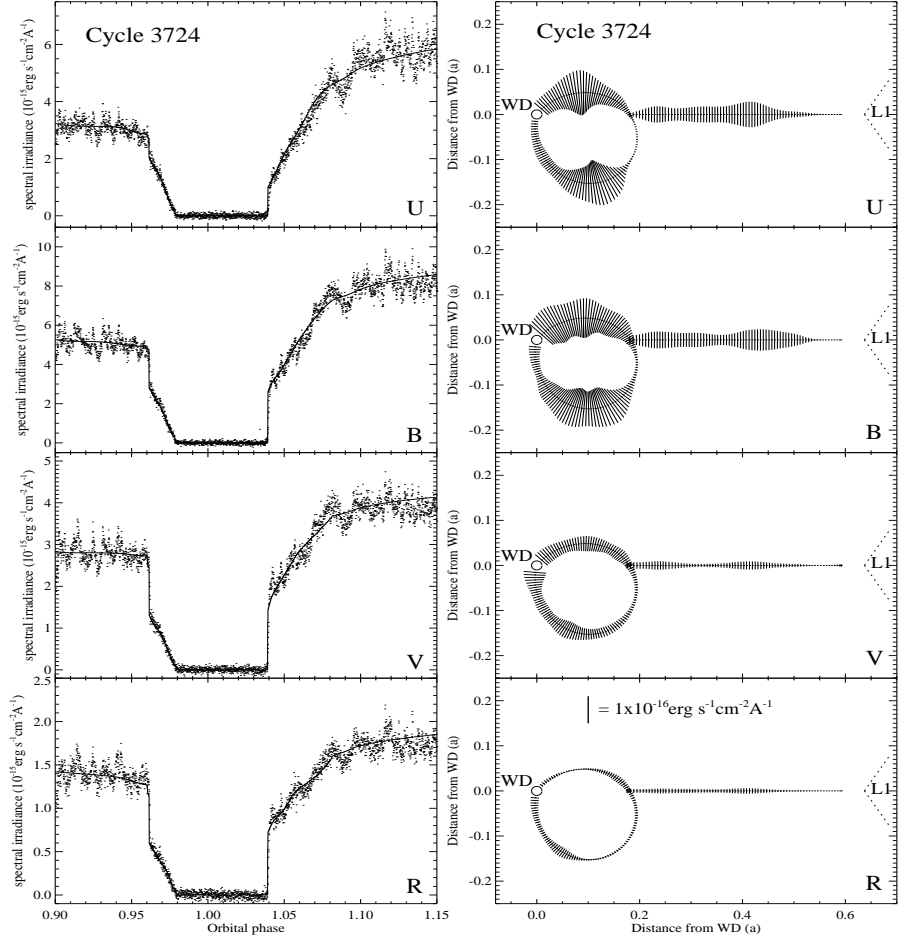,width=13cm,height=15cm}
\caption{Model eclipse profiles and the corresponding images of the accretion
stream for cycle 3724 (according to the ephemeris in Schwope, Mantel \& Horne
1997), using a stream that accretes onto both footpoints of a dipole field
line. The stream images are shown projected onto the plane perpendicular to the
orbital plane and passing through the centre of both stars. The brightness of
each emission point is shown as a line through the point, perpendicular to the
stream; the length of each line indicates the brightness of the point. The
white dwarf is shown to scale as a circle (labelled `WD'), and the secondary is
shown (not to scale) to mark the position of the L1 point (labelled `L1').}
\label{fig:cycle3724_D}
\end{figure}

Figure~\ref{fig:cycle3724_D} shows sample results for the two-footpoint
geometry model. Good fits to all five cycles were obtained using an orbital
inclination of $85^{\circ}$ and a mass ratio of 0.25. The ballistic stream is
diverted by the magnetic field at a radius $R_{\mu}$ in the range $0.21\,a \la
R_{\mu}\la 0.23\,a$ (where $a$ is the orbital separation of the two stars). The
magnetic colatitude $\beta$ of the dipole field is in the range
$22^{\circ}\la\beta\la 27^{\circ}$ and its longitude $\zeta$ is $10^{\circ}$
(this latter value is not well-constrained by the model; equally good fits were
obtained for values in the range 0--30$^{\circ}$). There are no obvious trends
in $\beta$ or $R_{\mu}$ over the three nights.

The stream images show that the brightness is not uniform along the stream, nor
is it a simple function of the radial distance from the white dwarf. Rather,
there are regions of the stream with localized beightness enhancements,
particularly within the magnetosphere. Interestingly, there is no obvious
brightening of the stream as it approaches either the threading radius or the
white dwarf.

\begin{figure}
\hspace*{-3mm}\psfig{file=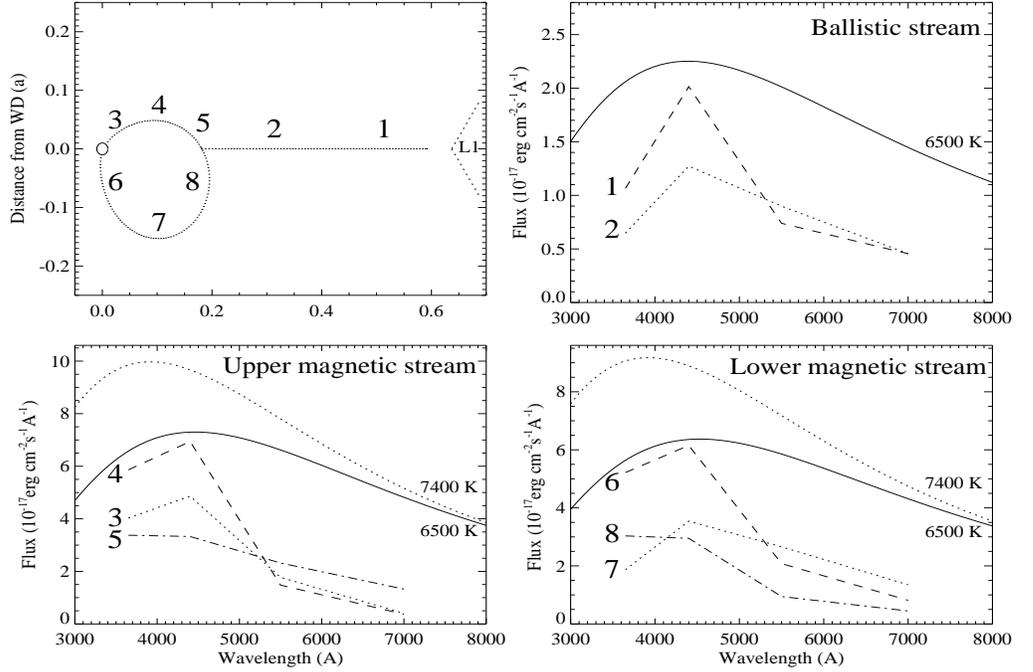,width=14cm,height=9.5cm,angle=90}
\caption{The colour-dependence of the stream brightness distributions for cycle
3723 (similar results are found for the other four cycles). The stream is
divided into eight sections (top left panel) and the average flux per stream
emission point in each section is plotted in the four wavebands for each of the
eight sections of the stream. Note the comparable $U$ and $B$ fluxes in
sections 5 and 8. Blackbody curves for temperatures of 6500\,K and 7400\,K are
shown for illustration.}
\label{fig:temperatures_D}
\end{figure}

To investigate the wavelength dependence of the stream images, the stream is
divided into eight sections and the average flux per stream emission point in
each section is examined in turn. In all stream sections apart from sections
`5' and `8' (see Fig.~\ref{fig:temperatures_D}), and in all cycles, the stream
flux is highest in $B$ and successively lower in $U$, $V$ and $R$. In four of
the five cycles, the stream flux is comparable in $U$ and $B$ in sections `5'
and `8'. 

\begin{figure}
\vspace*{-1cm}
\hspace*{1.7cm}\psfig{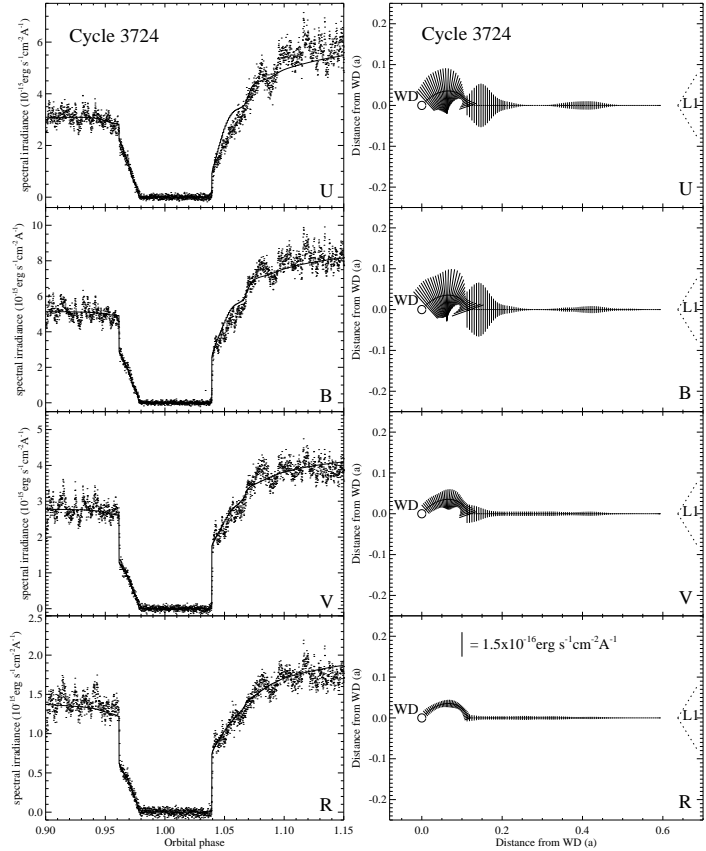}
\caption{As for Fig.~\ref{fig:cycle3724_D}, but for a model stream that 
accretes onto only the footpoint of the field line above the orbital plane (on
the same side of the orbital plane as the observer).}
\label{fig:cycle3724_S}
\end{figure}

Fig.~\ref{fig:cycle3724_S} shows the results for the same eclipse profile as in
Fig.~\ref{fig:cycle3724_D}, but modelled using a stream that accretes onto only
the pole on the same side of the orbital plane as the observer. The
one-footpoint model fits were obtained using $\beta=30^{\circ}$ and
$\zeta=10^{\circ}$ (once again $\zeta$ is not well-constrained). The coupling
radius decreases from 0.20\,$a$ to 0.17\,$a$ over the three nights. This
results in an increase in the angle between the magnetically-channeled part of
the stream and the line of centres, from $42^{\circ}$ to
$47^{\circ}$. Interestingly, this is almost identical to the range of angles
implied by the movement of the pre-eclipse dip.

The stream images are quite different from the two-footpoint geometry
results. There is very little emission from the ballistic stream. The stream
brightens as it approaches $R_{\mu}$. As the stream leaves the orbital plane it
fades, and then brightens again as it approaches the white dwarf. The brightest
part of the stream is not always closest to the white dwarf, though: in several
cycles the peak emission occurs about 0.08\,$a$ from the surface.

We examine the wavelength dependence of the stream brightness as before by
dividing the stream into sections. For this geometry we divided both the
ballistic and the magnetically-channeled part of the stream into two
sections. In all cycles and all stream sections (without exception) the flux is
highest in $B$, and is successively lower in $U$, $V$ and $R$.

\section{What can be deduced about the accretion stream?}
The range of values obtained for the coupling radius is 1.0--1.3$\times
10^{10}$\,cm, or 17--21 white dwarf radii. The radius of the stream can be
estimated using the pre-eclipse absorption dips; we find a radius of $3\times
10^{9}$\,cm. If we make the canonical assumption that the ram pressure of the
stream is balanced by the magnetic pressure at $R_{\mu}$, we can estimate the
mass transfer rate in the stream: this is found to be 8--76$\times
10^{16}$\,g$s^{-1}$. The smaller values in this range are in agreement with the
mass transfer rate obtained by Heerlein, Horne \& Schwope (1998) from their
models of the accretion flow in HU Aqr.

What emission mechanisms are operating in the stream to provide sufficient
optical flux to rival the flux from the accretion region? The stream flux is
not likely to be cyclotron emission due to its low levels of polarization (this
is can be seen from simultaneous intensity and polarization light
curves). Also, the cyclotron emission would be strongest where the thermal
velocity of electrons is highest. The bulk of the cyclotron radiation is thus
expected to be radiated from the post-shock flow close to the white dwarf
surface, and is unlikely to be the origin of the emission many white dwarf
radii from the white dwarf surface.

We can also eliminate line emission as the chief emission mechanism. This is
because line emission is strongest in the ballistic stream (as seen in
contemporaneous Doppler tomograms: Schwope et al.~1997), whereas the {\em
total} line and continuum emission (as seen in the stream images) is greatest
in the magnetosphere. Free-free emission, although providing sufficient flux,
does not have the correct spectral shape to match the wavelength dependence of
the stream.

As shown in Fig.~\ref{fig:temperatures_D}, the stream fluxes in {\em UBVR} are
closer to blackbody spectra. They do not match a blackbody curve exactly,
deviating in the $U$ and $R$ fluxes which fall significantly below a blackbody
curve. However, the stream fluxes in the various sections of the stream peak in
$B$ in most cases, and this is roughly consistent with a blackbody of
temperature 6500\,K. The exceptions to this occur in the case of the
two-footpoint geometry models, where the stream sections on the magnetic part
of the stream immediately adjacent to $R_{\mu}$ have $U$ fluxes comparable to
$B$. This suggests that the stream plasma that has just been threaded is hotter
than the fully-threaded plasma closer to the white dwarf and than the
free-falling plasma before the threading radius. This may be due to shock
heating of the plasma as it threads onto the magnetic field, as suggested by
e.g. Liebert \& Stockman (1985) and Hameury, King \& Lasota (1986).


\begin{references}
\reference Harrop-Allin M.K., Hakala P.J., Cropper M., 1998, \mnras, in press
\reference Heerlein C., Horne K., Schwope A.D., 1998, \mnras, in press
\reference Hameury J.-M., King A.R., Lasota J.-P., 1986, \mnras, 218, 695
\reference Liebert J., Stockman H.S., 1985, in Lamb D.Q., Patterson J., eds,
Cataclysmic Variables and Low-Mass X-ray Binaries. Reidl, Dordrecht, p.~151 
\reference Schwope A.D., Mantel K.-H., Horne K., 1997, A\&A, 319, 894

\end{references}
\end{document}